\documentstyle[aps,prl,epsf]{revtex}
\begin{document}
\twocolumn[\hsize\textwidth\columnwidth
\hsize\csname@twocolumnfalse\endcsname
\draft
\title{Magnetic Anisotropy of a Single Cobalt Nanoparticle}
\author{M. Jamet$^{1}$, W. Wernsdorfer$^{2}$, C.
Thirion$^{2}$,
D. Mailly$^{3}$, V. Dupuis$^{1}$, P. M\'elinon$^{1}$, 
A. P\'erez$^{1}$}
\address{
$^{1}$ D\'epartement de Physique des Mat\'eriaux,
        Universit\'e Claude Bernard-Lyon 1 et CNRS, 
	69622 Villeurbanne, France.\\
$^{2}$ Laboratoire Louis N\'eel, CNRS, BP 166, 
        38042 Grenoble Cedex 9, France.\\
$^{3}$ Laboratoire de Microstructures et de Micro\'electronique,
        96 av. H. Ravera, 92220 Bagneux, France}
%
\date{1 Aug. 2000}
\maketitle
\begin{abstract}
Using a new microSQUID set-up, we investigate magnetic anisotropy in a
single 1000-atoms cobalt cluster. This system opens new fields in the
characterization  and the understanding of the origin of magnetic
anisotropy in such nanoparticles. For this purpose, we report
three-dimensional switching field measurements performed on a 3 nm cobalt
cluster embedded in a niobium matrix. We are able to separate the
different magnetic anisotropy contributions and evidence the dominating
role of the cluster surface.
\end{abstract}
\vskip1pc]
\narrowtext
Magnetic nanostructures and nanomagnetism are subjects of growing interest
on account of the potential applications in the fields of high density
magnetic recording media and spin electronics. On the basis of the
increase in the average storage density observed in the past ten years
associated to a continuous decrease of the magnetic particle size, it has
been predicted that the superparamagnetic limit \cite{Dorm97} 
will be reached around
2005 with particle sizes around 10 nm  \cite{Sell99}. 
In order to overcome this
limit a better understanding of the magnetism in monodomain
particles is necessary. Because of the limited sensitivity of conventional
magnetic characterization techniques \cite{Wern00}, most of the
experimental studies
on nanosized grains were carried out on large assemblies of particles
\cite{Koda99}
where distributions of particle sizes, shapes and defects rendered the
interpretations quite difficult. Here we present the first magnetization
reversal measurement performed on individual cobalt clusters of 3 nm in
diameter (i.e. particles containing about one thousand atoms) prepared
with a low energy cluster beam deposition technique \cite{Pere97}. A new
microSQUID
set-up \cite{Wern00} measures the three dimensional diagram of the
magnetization
switching fields which is described with a uniform rotation model. We
deduce the magnetic anisotropies of such individual nanoparticles which
are dominated by surface anisotropy.\\
In bulk magnetic materials (3D), magnetostatic and bulk magnetocrystalline
energies are the main sources of anisotropy whereas in low dimensional
systems such as thin films (2D), wires (1D) or clusters (0D) strong
interfacial effects are expected \cite{Dorm97,Naka98,Dora98}. Only
experiments on a single
cluster can provide information on the different contributions to the
magnetic anisotropy. Here we present the first magnetic measurements on
individual cobalt clusters of 3 nm in diameter. High Resolution
Transmission Electron Microscopy (HRTEM) performed on Co clusters deposited on
carbon coated copper grids showed that they are well crystallized in the
f.c.c structure (Fig. 1a) with a sharp size
distribution (3-4 nm). Similar cobalt clusters are then embedded in a thin
niobium film for magnetic measurements and x-ray diffraction measurements
showed that embedded clusters keep their f.c.c structure. Clusters mainly
form truncated octahedrons\cite{Tuai97,Pare97}. 
\begin{figure}
\centerline{\epsfxsize=8cm \epsfbox{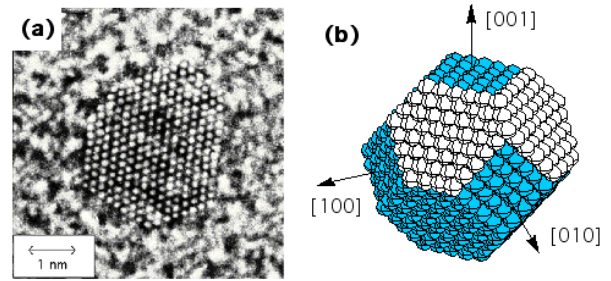}}
\caption{(a) High Resolution Transmission Electron Microscopy (HRTEM)
observation along a [110] direction of a typical 3 nm cobalt cluster
exhibiting a f.c.c structure. 
(b) A characteristic cluster simulated for our 
magnetic calculations with dark
atoms belonging to the 1289-atoms truncated 
octahedron basis and light
atoms to the (111) and (001) added facets.}
\label{FIG1}
\end{figure}
Faceting is thermodynamically
favorable to minimize the surface energy\cite{Butt83} leading to an
equilibrium
shape in the truncated octahedron form. Such perfect polyhedrons contain
1289 or 2406 atoms for diameters of 3.1 or 3.8 nm, respectively. As
previously observed for free and deposited metallic clusters (Co, Ni,
Al)\cite{Pell94,Mart92}, the growth of a polyhedron to one which is one
layer larger
occurs by the filling of successive facets. This result has also been
theoretically predicted using molecular dynamics\cite{Valk98}. In a first
stage,
atoms have a high probability ($\approx$80 $\%$) to participate to the
growth of a
close-packed (111) face, which will be the first covered. In a second
stage atoms will fill an adjacent (111) or (100) face (Fig. 1b). The magnetic
signals of such particles are at least a thousand times smaller than those of
previously studied nanoparticles \cite{Wern97,Bone99} 
deposited on the microSQUID device.
In order to achieve the needed sensitivity, 
Co-clusters preformed in the gas phase are directly
embedded in a co-deposited thin superconducting niobium 
film \cite{Jame00} which is
subsequently used to pattern microSQUID loops. 
A laser vaporization
and inert gas condensation source is used to produce an intense supersonic
beam of nanosized Co-clusters which can be deposited in various matrices
in UHV conditions. In such a low energy deposition regime (LECBD: Low
Energy Cluster Beam Deposition) clusters do not fragment upon impact on
the substrate \cite{Pere97}. The niobium matrix is simultaneously deposited
thanks
to a UHV electron gun evaporator. By monitoring both evaporation rates
using quartz balance monitors, it is possible to continuously adjust the
cluster concentration in the matrix. As prepared 20nm-thick niobium films
containing a very low concentration of cobalt clusters ($<$ 0.1 $\%$) are
electron beam lithographed to pattern micro-bridge-DC-SQUIDs of 1 $\mu$m in
dimension \cite{Wern95} (Fig. 2). The later ones allow us to detect the
magnetization reversal of a single Co-cluster for an applied magnetic
field in any direction and in the temperature range between 0.03 and 30 K.
However, the desired sensitivity is only achieved for Co-clusters embedded
into the micro-bridges where the magnetic flux coupling is high enough.
Due to the low concentration of embedded Co-clusters, we have a maximum of
5 non-interacting particles in a micro-bridge (300$\times$50 nm$^{2}$). We can
separately detect the magnetic signal for each cluster. Indeed they are
clearly different in intensity and orientation because of the random
distribution of the easy  magnetization directions.\\
For cobalt, the exchange length is 7 nm which is larger than the 3 nm
particle size \cite{Scha91}. In this case, we can use to a good
approximation the
Stoner and Wohlfarth model \cite{Ston91,Thia98} describing the
magnetization reversal
by uniform rotation. This model supposes that the exchange interaction in
the cluster couples all the spins strongly together to form a giant spin
which direction is described by the unit vector {\bf m}. The only degrees of
freedom of the particle magnetization are then the two angles
($\theta$,$\phi$) of the
orientation of {\bf m}. 
\begin{figure}
\centerline{\epsfxsize=5.5cm \epsfbox{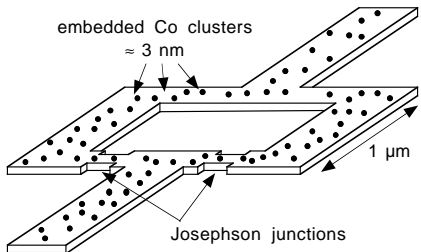}}
\caption{Schematic drawing of a micro-bridge-DC-SQUID which is patterned
out of a 20 nm-thick superconducting niobium film containing a low density
of 3 nm cobalt clusters (black dots). The concentration is low enough ($<$
0.1 $\%$) in order to have no more than 5 particles located in a micro-bridge
(300 $\times$ 50 nm$^{2}$). The magnetic flux coupling of only the clusters
in the
micro-bridges was strong enough to give a measurable signal for each
individual cluster. This new configuration detects the magnetization
reversal of few hundred of spins.}
\label{FIG2}
\end{figure}
\begin{figure}
\centerline{\epsfxsize=4.4cm \epsfbox{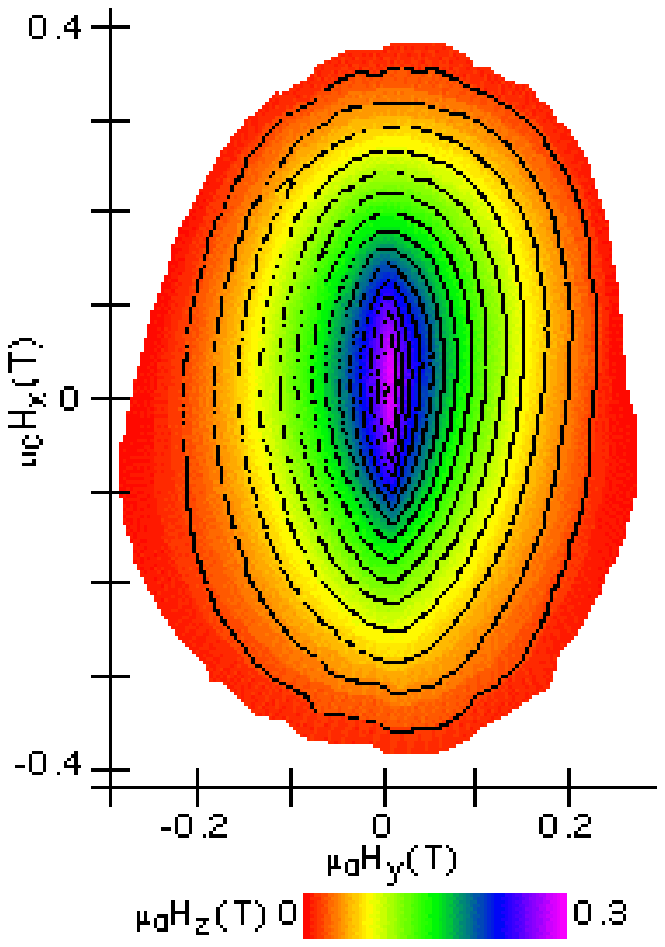}
            \epsfxsize=4.4cm \epsfbox{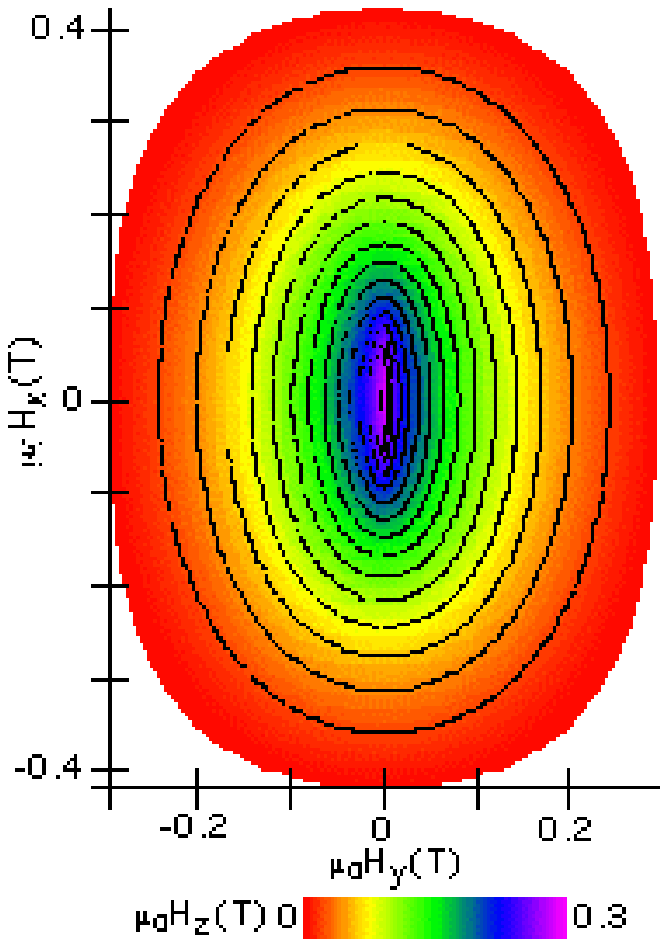}}
\centerline{\epsfxsize=7cm \epsfbox{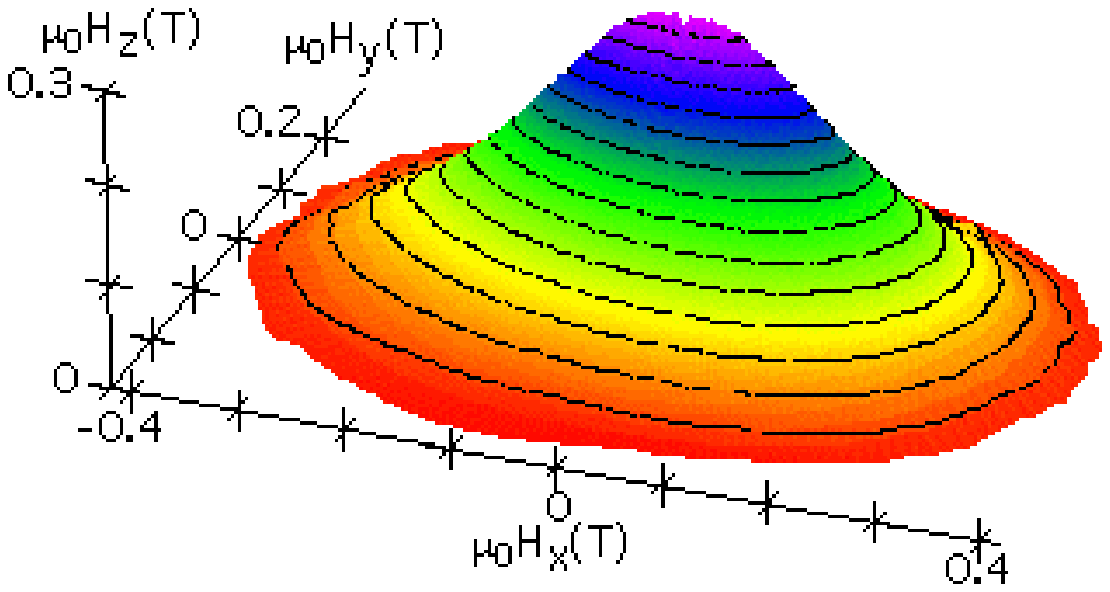}}
\centerline{\epsfxsize=7cm \epsfbox{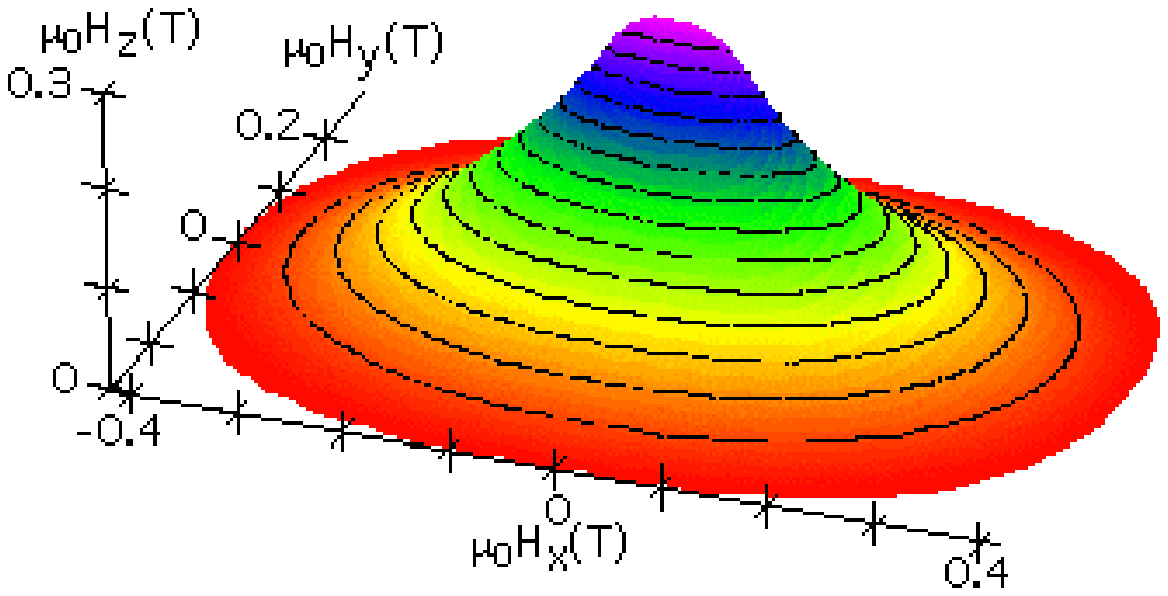}}
\caption{(color) (a), (c) Top view and side view respectively of the
experimental
three dimensional angular dependence of the switching field of a 3 nm
Co-cluster measured with the microSQUID. This surface is symmetrical with
respect to the H$_{x}$ - H$_{y}$ plane and only the upper part
($\mu_{0}$H$_{z} >$ 0 T) is
shown, it corresponds to almost 2000 measuring points. Continuous lines on
the surface are contour lines on which $\mu_{0}$H$_{z}$ is constant.
(b), (d) Top view and side view respectively of the
theoretical switching field surface considering only second and fourth
order terms in the anisotropy energy.}
\label{FIG3}
\end{figure}
The magnetization reversal is described by the
potential
energy:

\begin{equation}
E({\bf m},{\bf H})=E_{0}({\bf m})-\mu_{0}vM_{s}{\bf m}.{\bf H}
\end{equation}

where v and M$_{s}$ are the magnetic volume and the saturation
magnetization of
the particle respectively, {\bf H} is the external magnetic field.
E$_{0}$({\bf m}) is the magnetic anisotropy energy given by:

\begin{eqnarray}
E_{0}({\bf m})=E_{shape}({\bf m})+E_{surface}({\bf m})  \nonumber \\
+E_{ME}({\bf m})+E_{MC}({\bf m})
\end{eqnarray}

E$_{shape}$ is the magnetostatic energy related to the cluster shape.
E$_{surface}$ is due to the symmetry breaking and surface strains. In
addition, if the
particle experiences an external stress, the volume relaxation inside the
particle induces a magnetoelastic (ME) anisotropy energy: E$_{ME}$.
E$_{MC}$ is the
cubic magnetocrystalline anisotropy arising from the coupling of the
magnetization with the f.c.c crystalline lattice as in the bulk. All these
anisotropy energies can be developed in a power series of
m$_{x}^{a}$m$_{y}^{b}$m$_{z}^{c}$ with p $=$ a $+$ b $+$ c $=$ 2, 4, 6,...
giving
 the order of the anisotropy term. Shape anisotropy energy only contains
second order terms. Surface and magnetoelastic energies begin with second
order terms whereas the cubic
magnetocrystalline anisotropy starts with fourth order terms.\\
At T $=$ 0 K and {\bf H} $=$ {\bf 0}, {\bf m} is aligned along an easy
magnetization axis which
is a minimum of E. When a magnetic field {\bf H} is applied, the position
of the
minima in E changes continuously with {\bf m} following the position of a
minimum. However, there are particular fields where this minimum
disappears leading to a discontinuous variation of {\bf m} with a jump to
another minimum of E. The corresponding fields are called the switching
fields of the magnetization. The microSQUID technique measures the
switching fields for any direction of {\bf H} \cite{Bone99} allowing us to
determine the
magnetic anisotropy energy E$_{0}$ of a single cluster. The magnetization
switching is detected using the \textit{cold mode}\cite{Wern00}. In the
superconducting
state, the SQUID is biased close to the critical current. The
magnetization reversal of the particle then triggers the transition of the
SQUID to the normal state. The three dimensional switching field
measurements and the studies as a function of temperature were done using
a three step method (\textit{blind mode})\cite{Bone99}. First, the
magnetization of the
particles is saturated in a given direction (at T $=$ 35 mK). Then, a second
field is applied at a temperature between 35 mK and 30 K which may or may
not cause a magnetization switching. Finally, the SQUID is switched on (at
T $=$ 35 mK) and a field is applied in the SQUID plane to probe the
resulting magnetization state. This method allows us to scan the entire
field space. Fig. 3a and 3c display
a typical three dimensional switching field distribution for a 3 nm
Co-cluster at T $=$ 35 mK. This surface is directly related to the
anisotropy involved in the magnetization reversal of the particle.
The experimental results in Fig. 3a and 3c can be reasonably fitted with
the Stoner and Wohlfarth model \cite{Thia00} to obtain the following
anisotropy
energy:

\begin{eqnarray}
E_{0}({\bf
m})/v=-K_{1}m_{z}^{2}+K_{2}m_{y}^{2}-K_{4}(m_{x'}^{2}m_{y'}^{2} \nonumber \\
+m_{x'}^{2}m_{z'}^{2}+m_{y'}^{2}m_{z'}^{2})
\end{eqnarray}

K$_{1}$ and K$_{2}$ are the anisotropy constants along $z$ and $y$, the
easy and hard
magnetization axis respectively. K$_{4}$ is the fourth order anisotropy
constant and the ($x'y'z'$) coordinate system is deduced from ($xyz$) by a
45$^{\circ}$
rotation around the $z$ axis with $z'=z$. We find K$_{1} =$ 2.2 $\times$
10$^{5}$ J / m$^{3}$, K$_{2} =$
0.9 $\times$ 10$^{5}$ J / m$^{3}$ and K$_{4} =$ 0.1 $\times$ 10$^{5}$ J /
m$^{3}$. The corresponding theoretical
surface is showed in Fig. 3b and 3d. Furthermore, we measure the
temperature dependence of the switching field distribution (Fig. 4). We
deduce the blocking temperature of the particle T$_{B}$ $\approx$ 14 K
leading to an
estimation of the number of magnetic atoms in this particle: N $\approx$ 1500
atoms\cite{remark1}. Detailed measurements on about 20 different particles
showed similar three dimensional switching field distributions with
comparable anisotropy (K$_{1} =$ (2.0 $\pm$ 0.3) $\times$ 10$^{5}$ J /
m$^{3}$, K$_{2} =$ (0.8 $\pm$ 0.3) $\times$
10$^{5}$ J / m$^{3}$ and K$_{4} =$ (0.1 $\pm$ 0.05) $\times$ 10$^{5}$ J /
m$^{3}$)
and size (N $=$ 1500 $\pm$ 200
atoms).

\begin{figure}
\centerline{\epsfxsize=8cm \epsfbox{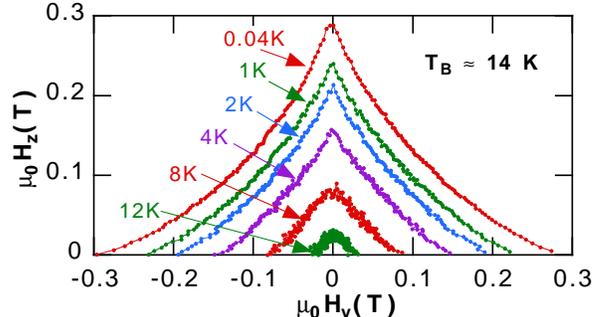}}
\caption{Temperature dependence of the switching field measured in the
H$_{y}$ -
H$_{z}$ plane in Fig. 3. An
extrapolation of the switching fields to zero gives the blocking
temperature T$_{B} =$ 14 K [22].}
\label{FIG4}
\end{figure}

In the following, we analyse various contributions to the anisotropy
energy of the small Co-clusters in view of the experimental results
reported above. Fine structural studies using EXAFS measurements
\cite{Jame00} were
performed on 500 nm thick niobium films containing a low concentration of
cobalt clusters ($<$ 5 $\%$). They showed that niobium atoms penetrate the
cluster surface to almost two atomic monolayers because cobalt and niobium
are miscible elements. Further magnetic measurements \cite{Jame00}  showed
that
these two atomic monolayers are magnetically dead. For this reason, we
estimate the shape anisotropy of the typical nearly spherical deposited
cluster in Fig. 1b after removing two atomic monolayers at the surface. By
calculating all the dipolar interactions inside the particle assuming a
bulk magnetic moment of $\mu_{at} =$ 1.7 $\mu_{B}$, we estimate the shape
anisotropy
constants: K$_{1} =$ 0.3 $\times$ 10$^{5}$ J / m$^{3}$ along the easy
magnetization
axis and K$_{2} =$ 0.1 $\times$ 10$^{5}$ J / m$^{3}$ along the hard
magnetization axis. These values are
much lower than the measured ones which means that E$_{shape}$ is not the main
cause of anisotropy in the cluster. We expect that the contribution of the
magnetoelastic anisotropy energy E$_{ME}$ coming from the matrix-induced
stress
on the particle is also small. Indeed, using the co-deposition technique,
niobium atoms cover uniformly the cobalt cluster creating an isotropic
distribution of stresses. In addition, they can relax preferably inside
the matrix and not in the particle volume since niobium is less rigid than
cobalt. We believe therefore that only interface anisotropy E$_{surface}$ can
account for the experimentally observed second order anisotropy terms.
Niobium atoms at the cluster surface might enhance this interface
anisotropy through surface strains and magnetoelastic coupling.
Quantitative information on surface anisotropy are only available in the
case of a cluster-vacuum interface using the N\'eel anisotropy model. This
phenomenological model is based on a simple atomic picture. In a
first approximation, the magnetic energy of a pair of atoms can be written
as: L cos$^{2}$($\theta$), where L is an atomic interaction and $\theta$
the angle between
the atomic bond and the magnetization. L depends on the f.c.c cobalt
magnetoelastic constants and at low temperature: L $=$ -1.5 $\times$
10$^{7}$ J / m$^{3}$
\cite{Chua94}. Summing over all the nearest neighbours in the f.c.c cobalt
cluster
in Fig. 1b, this interaction vanishes except at the cluster surface where
the cubic symmetry is broken. We have contributions from (111) and (100)
facets with in-plane anisotropy and from edges with an easy direction
along their axis. Apices give no contribution to the anisotropy since
locally the cubic symmetry is not broken. After removing two atomic dead
layers at the cluster surface, one finds: K$_{1} =$ 2.5$\times$10$^{5}$ J /
m$^{3}$ along the
easy direction and K$_{2} =$ 0.5$\times$10$^{5}$ J / m$^{3}$ along the hard
magnetization axis.
Therefore, the N\'eel surface anisotropy  involves very large anisotropy
constants in thin films (10$^{7}$ J / m$^{3}$) whereas in clusters, symmetries
reduce this anisotropy to a value close to our experimental result.
The fourth order term K$_{4} =$ 0.1 $\times$ 10$^{5}$ J / m$^{3}$ gives the
cubic
magnetocrystalline anisotropy in the f.c.c cobalt cluster. $x'$, $y'$ and
$z'$ correspond to the crystallographic directions [100], [010] and [001]
respectively thus [111] directions are weak easy magnetization axes (Fig. 3).
The anisotropy constant is smaller than the one reported in previous works
\cite{Chua94,Lee90}.
But, in our case, surface atoms which atomic environment may deviate from
the pure f.c.c one give a large contribution to this magnetocrystalline
anisotropy.\\
In conclusion, we have shown that the microSQUID technique combined with
the Low Energy Cluster Beam Deposition is a powerful method to investigate
the magnetic properties of nanosized magnetic particles. In particular, it
allows to measure in three dimensions the switching field of individual
grains giving access to its magnetic anisotropy energy. Furthermore, the
temperature dependence of the switching field is measurable and allows to
probe the magnetization dynamics. In the case of nanosized cobalt clusters
embedded in the niobium film of the microSQUID, it seems that the
cluster-matrix interface provides the main contribution to the magnetic
anisotropy. Such interfacial effects could be promising to control the
magnetic anisotropy in small particles in order to increase their blocking
temperature up to the required range for applications.\\
The authors are indebted to B. Pannetier, A. Benoit, 
F. Balestro and J.P. Nozi\`eres
from CNRS-Grenoble for their contributions to the improved microSQUID
technology and to G. Guiraud from DPM-Villeurbanne for his contribution to
the cluster deposition. Part of this work has been supported by DRET,
Rh\^one-Alpes, and the MASSDOTS ESPRIT LTR-Project $\#$32464.

\newpage
\widetext
\end{document}